\documentclass[%
 aps,
 prd,
 twocolumn,
 nofootinbib,
 floatfix,
 reprint
]{revtex4-2}

\usepackage{amsmath}
\usepackage{amssymb}
\usepackage{mathrsfs} 

\usepackage{graphicx} 
\usepackage[caption=false]{subfig} 

\usepackage{hyperref} 
\usepackage[svgnames]{xcolor} 

\usepackage{booktabs} 
\usepackage{dcolumn} 

\usepackage{physics}
\usepackage{comment}

\definecolor{FinnishBlue}{rgb}{0,0.435,1}

\newcommand{\sectioninline}[1]{\textit{#1}---} 
\newcommand\cb[0]{\phi_{\text{cb}}}
\newcommand\meta[0]{\phi_{0}}
\newcommand{\Nsites}{N_\text{sites}}

\begin{document}

\title{Langer's nucleation rate reproduced on the lattice}

\author{Joonas Hirvonen}
\email{joonas.hirvonen@nottingham.ac.uk}

\author{Oliver Gould}
\email{oliver.gould@nottingham.ac.uk}

\affiliation{
    School of Physics and Astronomy,
    University of Nottingham,
    Nottingham NG7 2RD,
    U.K.
}

\date{May 28, 2025}

\begin{abstract}
We show that Langer's rate of bubble nucleation is quantitatively correct up to small higher-loop corrections, in comparison to lattice simulations.
These results are a significant advancement on decades of lattice studies showing only qualitative trends, and the first showing agreement for any conservative system.
We confirm that the failure to fully thermalize the metastable phase explains discrepancies with recent lattice studies that found disagreement with Langer's rate.
The key theoretical development is the translation of Langer's perturbative definition of a thermal metastable phase into a nonperturbative statement that can be implemented on the lattice.
Our statistical and systematic errors are small enough to allow us to measure on the lattice the coefficient of the two-loop contribution, missing from the perturbative prediction.
Our conclusions also exclude a possible systematic uncertainty in $^3$He experiments.
\end{abstract}

\maketitle

\sectioninline{Introduction}
The study of bubble nucleation in homogeneous systems goes back at least 150 years to Gibbs~\cite{Gibbs1874}, who introduced the notion of a critical bubble, for which the surface tension balances against the outward pressure.
Important early theoretical developments in this field include Becker and D\"{o}ring's discrete model for the dynamics of growing bubbles~\cite{becker1935kinetische},
Wigner's application of the transition state idea to chemical reactions~\cite{wigner1938transition},
and Zeldovich's computation of the nucleation rate by modeling the bubble radius as a particle undergoing Brownian motion~\cite{zeldovich1942theory}.
This latter result borrowed heavily from earlier work by Kramers on the escape rate of a classical thermal particle from a deep potential well~\cite{Kramers:1940zz}.
A field theoretic description of critical bubbles was introduced by Cahn and Hilliard~\cite{cahn1959free3}.

The modern theory of homogeneous nucleation was initiated by Langer in the late 1960s~\cite{Langer:1967ax, Langer:1969bc}.
This constitutes a dynamical framework for studying nucleation, based on a description of the system in terms of stochastic Hamilton's equations.
Langer's work is a generalisation of Kramer's analysis of a particle's escape rate to an arbitrary number of degrees of freedom and to field theory~\cite{Berera:2019uyp}.
This is sufficiently flexible to describe a wide range of different systems.

For instance, consider a system in $d$ spatial dimensions described by a scalar field $\phi(t, x)$, canonical momentum field $\pi(t, x)$, and the following Hamiltonian equations
\begin{align}
    \partial_t \phi &= \pi \;, &
    \partial_t \pi &= -\frac{\delta H}{\delta \phi} \,,
\end{align}
where $H$ is the effective Hamiltonian.
The system is thermalized locally around a homogeneous metastable background $\meta$, normalized such that $H[\meta]\equiv H[\phi=\meta, \pi=0]=0$.
The temperature is $T=1/\beta$.

The rate per unit volume at which bubbles form was found to be~\cite{Langer:1967ax, Langer:1969bc}
\begin{equation} \label{eq:langer_rate}
    \Gamma_\text{Langer} = 
    \frac{V}{2\pi}
    \left(\frac{\beta H[\cb]}{2\pi}\right)^{\frac{d}{2}}
    \bigg|\frac{\det H^{(2)}[\meta]}{\det^{+} H^{(2)}[\cb]}\bigg|^{\frac{1}{2}}
    \ e^{-\beta H[\cb]}
\end{equation}
in terms of the critical bubble $\cb$, a saddlepoint of the effective Hamiltonian.
The superscript in $H^{(2)}$ denotes the second functional derivative, and the plus on $\det^{+}$ means that only positive eigenvalues are included.
Eq.~\eqref{eq:langer_rate} relies on the saddlepoint approximation to a path integral, valid if the rate is slow, and the system is weakly coupled.

Eq.~\eqref{eq:langer_rate} was a major theoretical advance because it was the first complete expression of the nucleation rate to be derived from the underlying equations that describe the system and its fluctuations.
As a consequence, the elements of Eq.~\eqref{eq:langer_rate} can all be computed self-consistently in terms of parameters in the Hamiltonian.
While it relies on a saddlepoint approximation, this can be systematically improved~\cite{Ekstedt:2022tqk}.
More recently, Refs.~\cite{Gould:2021ccf, Hirvonen:2024rfg} outlined the theory for matching to Eq.~\eqref{eq:langer_rate} from an underlying quantum field theory, expressing it in terms of the Lagrangian parameters of the microscopic theory.

Despite this provenance, and despite decades of effort, Langer's result has yet to be quantitatively tested.
Experimental setups studying water, alcohols, alcanes and other small molecules have in some cases found qualitative agreement on the nucleation rate, and in others disagreement by many orders of magnitude, including on the temperature dependence~\cite{oxtoby1992homogeneous}.
However, for such studies, the comparison to theory is obscured by uncertainties in modeling the physical system and in determining its microphysical parameters; see for instance Ref.~\cite{langer1980kinetics, nellas2010exploring}.
Further, the possibility of impurities seeding nucleation raises persistent concerns.

A long-running experimental program on the A-B transition in liquid $^3\text{He}$ has sought to overcome these challenges, using smooth walls or electromagnetic trapping to avoid contact with vessel walls, but unfortunately remains inconclusive~\cite{QUEST-DMC:2024crp}.
Recently, several experiments to test nucleation theory have been proposed which use analog systems~\cite{Fialko:2014xba, Fialko:2016ggg, Billam:2021nbc, Tian:2022dzv, Jenkins:2023eez, Jenkins:2023npg, Darbha:2024srr}.
Two such experiments have been performed~\cite{Song:2021pyy, Zenesini:2023afv}, showing qualitative agreement between theory and experiment, though this required scaling and fitting, rather than direct comparison.
Nevertheless, ongoing experimental efforts in several groups offer much promise.

In the absence of a quantitative experimental test, many researchers have investigated lattice simulations as a means to test homogeneous nucleation theory.
The first exploratory lattice studies were carried out in the late 1980s, and established the two main approaches to simulating the dynamics of thermal field theories.
Grigoriev and Rubakov included thermal fluctuations in the initial time slice and then evolved them under conservative Hamilton's equations~\cite{Grigoriev:1988bd}, while Bochkarev and de Forcrand simulated a Langevin equation, thereby sourcing thermal fluctuations throughout the evolution~\cite{Bochkarev:1989tk}.
Investigating kink-antikink production, both found a Boltzmann-like exponential dependence on temperature, however Ref.~\cite{Bochkarev:1989tk}, and later Ref.~\cite{Alford:1991qg}, found that the value of the exponent was significantly smaller than the semiclassical prediction. 

Extending this to nucleation from a metastable phase, Ref.~\cite{valls1990nucleation} found semiclassical and lattice results strongly disagreed on the nucleation rate in 2+1 dimensions.
Ref.~\cite{Borsanyi:2000ua} improved agreement to within a factor of $e^{O(10)}$ in the rate, using a loop-corrected effective potential.
Alford, Feldman and Gleiser investigated nucleation in both 1+1 and 2+1 dimensions~\cite{Alford:1993zf, Alford:1993ph} and were able to fit the semiclassical form of the logarithm of the rate to the lattice results within statistical errors, by treating the prefactor as a constant nuisance parameter.

Moore, Rummukainen and Tranberg developed an alternative lattice approach for calculating the nucleation rate in very slow transitions, based on a split of the problem into statistical and dynamical parts~\cite{Moore:2000jw, Moore:2001vf}.
Comparisons of the results of this method to perturbative predictions have shown qualitative agreement at best, with quantitative disagreement on the rate at the level of $e^{O(10)}$ to $e^{O(100)}$ in some cases~\cite{Moore:2000jw, Moore:2001vf, Gould:2022ran}.

In all these studies, the functional determinant in Eq.~\eqref{eq:langer_rate} was treated as an unknown fit parameter, or estimated based on simple dimensional arguments. However, recent efforts to compute functional determinants have made it possible to directly compute Langer's complete formula~\cite{Ekstedt:2021kyx, Ekstedt:2023sqc, Pirvu:2024nbe, Pirvu:2024ova}.
Including the functional determinant was found to significantly improve agreement in 3+1 dimensions~\cite{Gould:2024chm}, from $\sim e^{50}$ to $\sim e^{10}$.

By far the closest quantitative agreement between lattice simulations and Langer's formula was found in Refs.~\cite{Pirvu:2024nbe, Pirvu:2024ova, Shkerin:2025hui}. They considered the following scalar field model in 1+1 dimensions
\begin{align} \label{eq:model_action}
    H = \int \mathrm{d}x \left(\frac{\pi^2}{2}+\frac{(\partial_x\phi)^2}{2} + \frac{m^2\phi^2}{2} - \frac{\lambda\phi^4}{4}\right) \,.
\end{align}
The metastable phase in this model is at $\phi_0=0$, and nucleation leads to escape to $\phi\to \pm\infty$. All $m$ and $\lambda$ dependence can be factored out by scaling $x\to (1/m)x $, $\phi\to (m/\sqrt{\lambda}) \phi$ and $\pi\to (m^2/\sqrt{\lambda})\pi$. The sole remaining parameter which characterizes the system is then $\hat{T}\equiv \lambda T/m^3$.

Refs.~\cite{Pirvu:2024nbe, Pirvu:2024ova, Shkerin:2025hui} found Eq.~\eqref{eq:langer_rate}
to overestimate the measured rate by a factor of $\approx 8$.
The authors argued that this overestimate was due to a failure of the metastable phase to thermalise.
They considered a modified system with damping and noise terms to aid thermalization.
They found that, when these terms were sufficiently large, the lattice results agreed well with Langer's generalized rate formula, which extends Eq.~\eqref{eq:langer_rate} to include damping.
They also noted that the measured rate was time dependent, something previously observed in Ref.~\cite{Batini:2023zpi}.
This implied a fundamental ambiguity in the definition of the nucleation rate.

We take the conservative Hamiltonian model and parameter points of Refs.~\cite{Pirvu:2024nbe, Pirvu:2024ova} as our starting point, aiming to decisively test Langer's formula by resolving the ambiguity in the definition of the nucleation rate.

\sectioninline{Nucleation}
The first step to understanding or computing the nucleation rate is defining it.
While this might seem unambiguous, as we will argue below, some quantitative discrepancies in the literature can be understood as resulting from different definitions of the nucleation rate.

We start from the observation that nucleation from a metastable phase is by assumption slow.
It is relevant when the initial conditions are stable on short, microscopic timescales $t_\text{micro}$, but unstable on longer timescales, $t_\text{global}$, over which the field explores the full global potential and its multiple phases.
Note that in our model with unbounded potential, escaping trajectories do not return so $t_\text{global}$ is infinite.
A natural microscopic timescale is that over which a slightly super-critical bubble grows, $1/\sqrt{|\lambda_-|}$, in terms of the single negative eigenvalue $\lambda_-$ of fluctuations about the critical bubble.

\begin{figure}
  \includegraphics*[width=0.75\columnwidth]{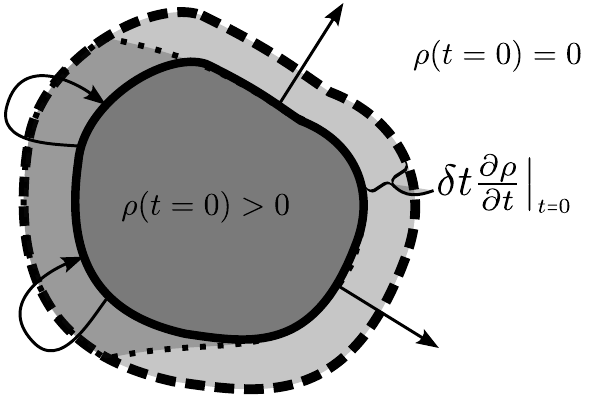}
  \caption{Growth in phase space of an initially localized distribution.
  The escape rate is defined as the initial rate of change of the distribution, excluding regions which soon return.
  The dark gray region bounded by the full line is the initial distribution.
  The region between the full line and the dashed line becomes occupied after a short time $\delta t>0$.
  Of this region, the medium gray region bounded by the dotted line rapidly returns.
  Only the light gray region, which leads to long-term escape, contributes to the escape rate.}
  \label{fig:phase_space_growing}
\end{figure}

\textit{Definition:} The \textit{escape rate} is the initial
escape rate
out of a region $\mathcal{R}$ in phase space, counting only those trajectories which do not return to $\mathcal{R}$ within any microscopic timescale.
The initial probability density $\rho(t=0)$ is taken to be nonzero within $\mathcal{R}$ and zero outside it; see Fig.~\ref{fig:phase_space_growing}.
As an equation, this is
\begin{equation}
    \Gamma \equiv
    -\int_\mathcal{R}  \mathcal{D}\phi \mathcal{D}\pi\
    \delta_\text{NR}(\mathcal{T})
    \frac{\partial \rho}{\partial t}\bigg|_{t=0} \,,
    \label{eq:rate_definition}
\end{equation}
where $\mathcal{D}\phi \mathcal{D}\pi$ is the phase space integration measure.
We have introduced a \textit{no-return} (NR) projector, $\delta_\text{NR}(\mathcal{T})$, which removes configurations that return to $\mathcal{R}$ within a time $\mathcal{T}$. This can be defined as
\begin{equation}
    \delta_\text{NR}(\mathcal{T}) \equiv \prod_{t=0}^\mathcal{T} \theta( \{\phi(t),\pi(t)\}\notin \mathcal{R}) \,,
\end{equation}
in terms of $\theta(\cdot)$, a Boolean step function which evaluates to 1 if the argument is true and 0 otherwise.
The no-return projector ensures we only count the last escape of each evolution trajectory, and so avoid double counting.
The timescale $\mathcal{T}$ is chosen such that $t_\text{micro} \ll \mathcal{T} \ll t_\text{global}$.
An alternative approach would be to count all crossings, and then divide by the number of crossings~\cite{Moore:2000jw, Moore:2001vf}.

The \textit{thermal nucleation rate} is the escape rate where $\mathcal{R}$ is identified with a metastable phase ($\text{meta}$) which is thermally occupied. So, the initial condition for the probability distribution is ${\rho(t=0)=\rho_\text{meta}}$, where
\begin{align}
    \rho_\text{meta}
    &\equiv
    \frac{
        e^{-\beta H[\phi, \pi]}
        \theta\left(\phi\in\text{meta}\right)
    }{Z_{\text{meta}}} \,,
    \label{eq:momentumHeatBathMetastableThermaDistributionPi}
\end{align}
and $Z_{\text{meta}}$ normalizes the distribution to unity.
This definition defines the metastable phase at a given instant.

The reasoning behind defining the nucleation rate at the initial time is as follows.
Over sufficiently long times, the system approaches global thermal equilibrium, where all net probability fluxes vanish.
If we were to define a rate as a function of time, it would inevitably vary with time and approach zero as $t\to \infty$.
To extract a well-defined and meaningful value for the nucleation rate, we must define it at a given time, or over a given time interval, for an appropriate initial distribution.

The exponential growth of super-critical bubbles ensures that bubbles which have remained outside the metastable phase for a time $\mathcal{T}$ are extremely unlikely to return in the near future, so that the rate should be exponentially insensitive to the precise choice of $\mathcal{T}$.

The next question arising is how to define the metastable phase, i.e.\ the set of metastable configurations $\phi\in \text{meta}$.
In mean field theory, the metastable phase is given by the value of some order parameter at a local minimum of the free energy, for instance $\phi = \meta$.
Perturbatively one can extend this definition to include small fluctuations.
However, for lattice simulations a nonperturbative definition is needed, which allows to unambiguously identify any given lattice configuration as either \textit{in} or \textit{not in} the metastable phase.
We propose the following.

\textit{Definition:} The \textit{metastable phase} is the set of field configurations that flow to the metastable minimum under gradient descent.
That is, $\phi(\tau\to\infty)\to\phi_0$ under
\begin{equation} \label{eq:gradient_descent}
    \frac{\mathrm{d}\phi}{\mathrm{d}\tau} = -\grad_\phi H[\phi,0] \,.
\end{equation}
This definition is nonperturbative, but nevertheless connects to mean field theory through the endpoint of the evolution.
The boundary of the metastable phase is a separatrix for Eq.~\eqref{eq:gradient_descent}.

The phase separation in a first-order phase transition comes from the potential energy barrier between phases. Applying gradient descent to a configuration takes it down the potential barrier either to the metastable phase, or not. A similar approach has been successfully utilized in nonperturbative computations of the sphaleron rate~\cite{Khlebnikov:1988sr, Moore:1998swa}.

This definition of the metastable phase is well-suited to the 1+1 dimensional model considered here, for which ultraviolet fluctuations yield only perturbatively small, finite contributions to $\delta H/\delta \phi$ in the continuum limit; see Appendix~\ref{app:continuumLimit}. However, Eq.~\eqref{eq:gradient_descent} flows to the minima of the bare potential so, in models where ultraviolet fluctuations are more important, this equation should be extended with a noise term, fixed by the fluctuation dissipation relation.
While this would make the identification of the metastable phase statistical, such fluctuations could be incorporated in the statistical uncertainty of the final nucleation rate.

Alternative definitions of the metastable phase are possible.
One could demand that the negative eigenmodes
of the field configurations, $\phi\in\text{meta}$, are not being pushed to grow by the Hamiltonian gradient.
This definition is related to that used for the nucleation moment in Ref.~\cite{Pirvu:2024nbe}, which looked (statistically) at the lowest point in the energy of the infrared modes of the conjugate momentum.

In Refs.~\cite{Langer:1967ax, Langer:1969bc}, Langer defined the nucleation rate similarly, as the probability flux out of the metastable phase. 
However, there are differences, which we expect to manifest in the rate at subleading orders.
First, the factor $\delta_\text{NR}(\mathcal{T})$ is omitted, because multiple crossings of the separatrix do not occur in the perturbative approach at Langer's order.
Further, the boundary conditions of the problem are modified to make the problem time independent, and hence analytically tractable.
This is done by introducing a source on the metastable side and a sink on the stable side of the separatrix to precisely compensate for the probability flux due to nucleation.

\sectioninline{Thermalizing the metastable phase}
To sample the metastable phase, we combine a Hamiltonian (or Hybrid) Monte Carlo~\cite{Duane:1987de} update with an accept/reject step for the condition $\phi \in \text{meta}$.
For further details and a proof of detailed-balance for the algorithm, see Appendix~\ref{app:Thermalization}.

The configuration $\phi$ is in the metastable phase if gradient descent takes the configuration asymptotically to zero, $\phi(\tau\to\infty)\to 0$. This happens exponentially quickly, with the exponent given by $|\lambda_-| \tau=3\tau$, so in practice only a relatively short period of gradient descent is needed.

For comparison, we consider a perturbative definition of the metastable phase: $\phi(t=0, x) = \phi_0 + \delta \phi(t=0, x)$, where $\phi_0=0$ is the mean-field metastable phase, and the fluctuations are Gaussian,
\begin{align}
    &\rho^\text{(G)}_\text{meta} = 
    \frac{
        e^{-\hat{\beta}\int \mathrm{d}x \big(\frac{\pi^2}{2}+\frac{(\partial_x\delta\phi)^2}{2} + \frac{m_T^2\delta\phi^2}{2}\big)}
        \theta(\phi \in \text{meta})
    }{Z^\text{(G)}_\text{meta}}
    \,.
\end{align}
Here $\hat{\beta}=1/\hat{T}$ and we have used the thermally corrected mass ${m_T^2 = 1 - \frac{3}{2}\hat{T}}$.
As before, configurations which are outside the metastable phase of the full potential are removed from the initial ensemble.
This accurately describes small fluctuations around the metastable minimum; see Appendix~\ref{app:numericalTests}.
However, it significantly under represents larger fluctuations.

\begin{figure}
  \includegraphics*[width=0.95\columnwidth]{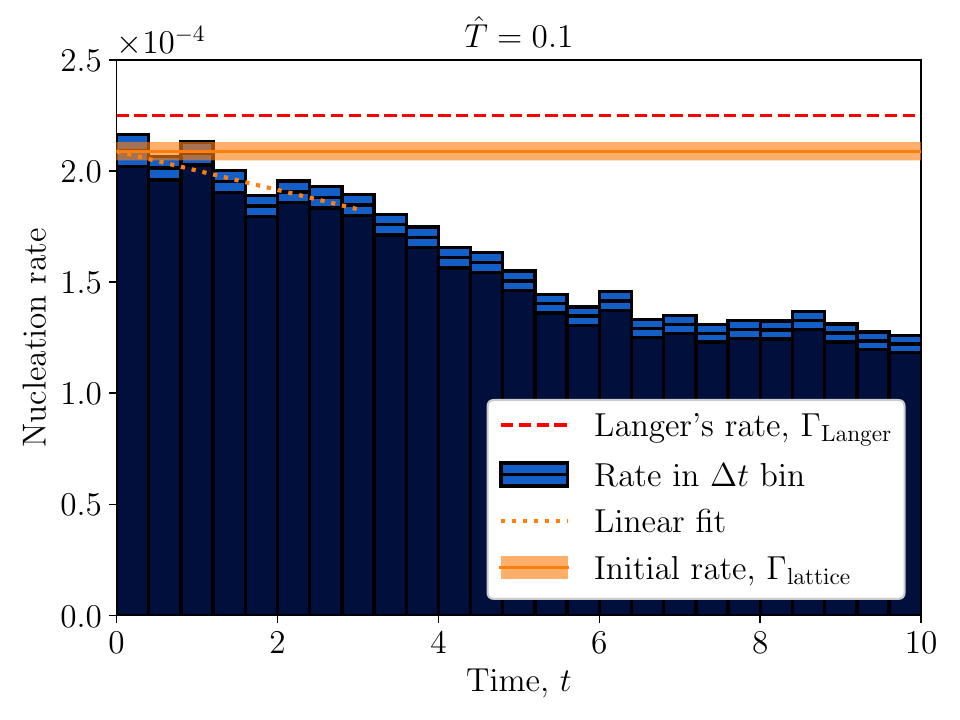}
  \caption{Nucleation rate at $\hat{T}=0.1$ for early times after full thermalization of the metastable phase at time $t=0$. Shown also is a linear extrapolation to $t\to 0$, which yields our initial rate and error. Note the good agreement with Langer's rate up to $\sim 10\%$, the expected size of two-loop terms missing in the perturbative computation. We use units where $m=1$.}
  \label{fig:rate_histogram_T0.1}
\end{figure}
\begin{figure}
    \includegraphics*[width=0.95\columnwidth]{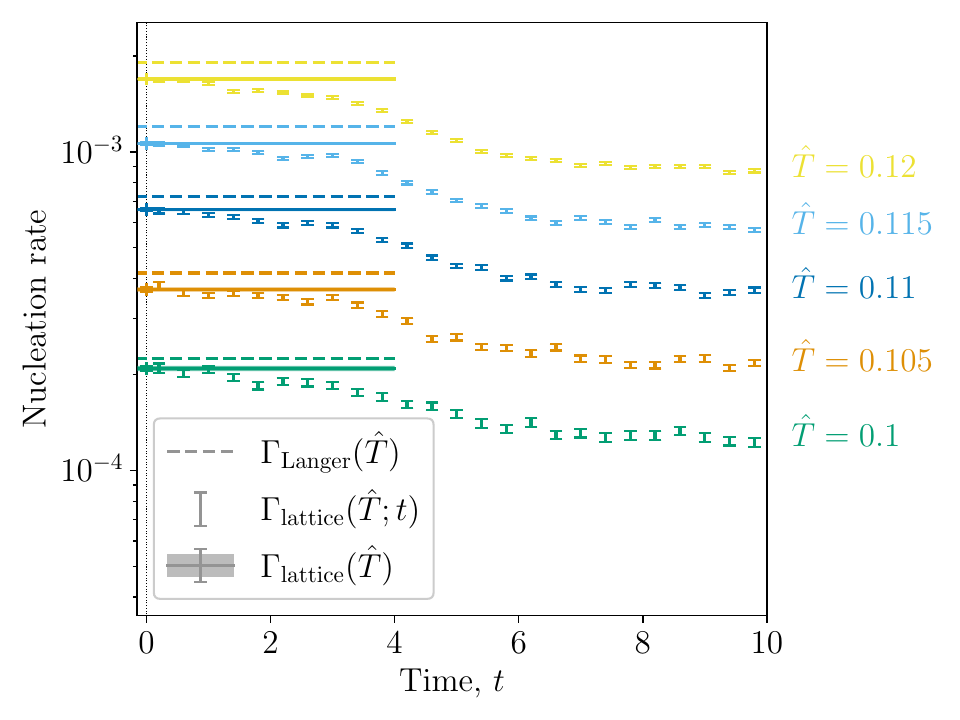}
    \caption{Nucleation rates across different temperatures, showing numerically computed initial rates together with Langer's formula. Data points $\Gamma_\text{lattice}(\hat{T};t)$ are the midpoints of the corresponding histograms, and the bands labeled $\Gamma_\text{lattice}(\hat{T})$ are the results of linear extrapolation to $t\to 0$ as in Fig.~\ref{fig:rate_histogram_T0.1}.}
    \label{fig:rates_compared}
\end{figure}
\begin{figure}
  \includegraphics*[width=0.95\columnwidth]{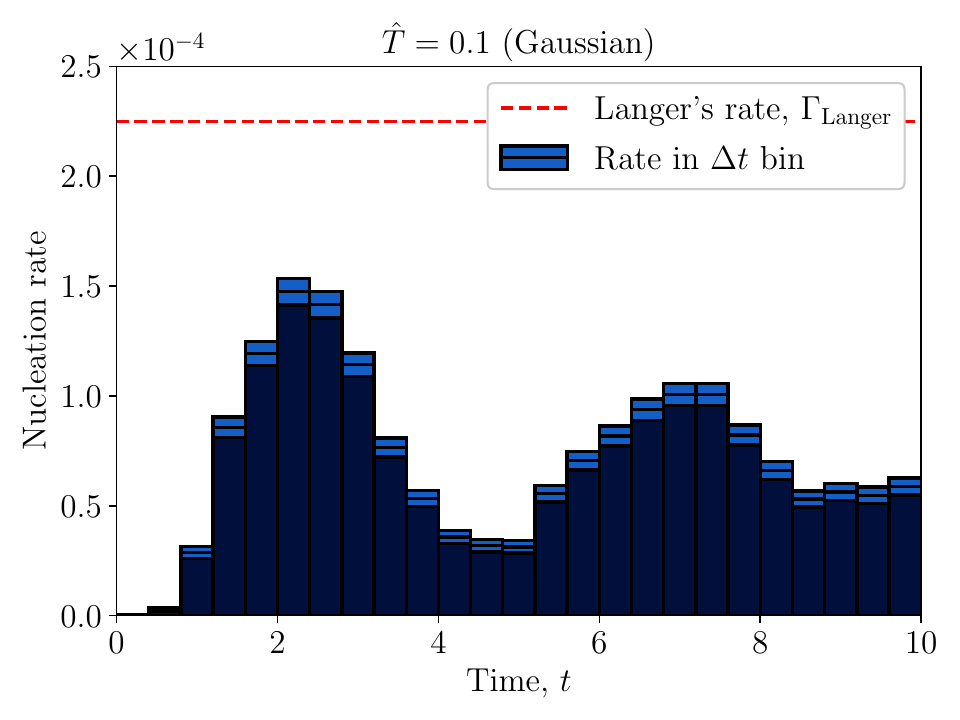}
  \caption{Nucleation rate at $\hat{T}=0.1$ for early times after Gaussian initialization around metastable phase minimum at time $t=0$.}
  \label{fig:rate_histogram_gaussian_T0.1}
\end{figure}

\sectioninline{Results}
Figs.~\ref{fig:rate_histogram_T0.1} and \ref{fig:rates_compared} summarise our main results, and Fig.~\ref{fig:rate_histogram_gaussian_T0.1} shows a comparison with Gaussian initial fluctuations.

We have performed simulations for five temperatures $\hat{T}\in \{0.1, 0.105, 0.11, 0.115, 0.12\}$. In units where $m=1$, we chose lattice extent $L=50\gg 1$, lattice spacing $\delta x=0.1\ll 1$ and time step $\delta t = 0.02 < \delta x$.
We ensured that our results are independent of the precise choice of lattice parameters; see Appendix~\ref{app:numericalTests}.

For each $\hat{T}$, we produced $N=2\cdot10^7$ independent configurations in the metastable phase.
Our final results use the gradient descent definition, but an alternative definition based on the eigenmodes of $\delta^2 H/\delta \phi^2$ gave consistent results.

The no-return criterion $\delta_\text{NR}(\mathcal{T})$ was implemented by evolving each configuration to a time $\mathcal{T}=10\gg 1/\sqrt{|\lambda_-|}=1/\sqrt{3}$, under conservative Hamiltonian dynamics.
If it is outside the metastable phase, we backtrack to the last step in the metastable phase, $t_{\text{last meta}}$ and define $t_\text{escape}=t_{\text{last meta}}+\delta t/2$ as the nucleation event.
A configuration contributes to $N_\text{meta}(t)$ for $t<t_\text{escape}$.

We binned $N_\text{meta}(t)$ in time, choosing bin widths $\Delta t$ via Scott's rule~\cite{binning}, and approximated the nucleation rate by mid-point differentiation $\Gamma_\text{lattice}(\hat{T}; t + \Delta t/2) = -\left[N_\text{meta}(t+\Delta t) - N_\text{meta}(t)\right]/\Delta t$.

Evaluating the critical bubble and functional determinant in this model~\cite{Pirvu:2024nbe}, Langer's rate reads
\begin{equation} \label{eq:langer_in_our_model}
    \Gamma_\text{Langer}(\hat{T}) =
    L \frac{6}{\pi}\sqrt{\frac{2}{3\pi \hat{T}}}\ e^{-4/(3\hat{T})}
    \,,
\end{equation}
in terms of the microscopic parameters.
Subleading $l\geq 2$ loop corrections are suppressed relatively by $c_{l}\hat{T}^{l-1}$, with each $c_{l}=O(1)$.

Fig.~\ref{fig:rate_histogram_T0.1} shows the nucleation rate at $\hat{T}=0.1$. A linear fit to $t\in[0, 3]$ and extrapolation to $t \to 0$ yields $\Gamma_\text{lattice}(\hat{T})$; a quadratic fit agrees within errors.
The lattice result (orange band) is shown alongside Langer's rate (dashed line).
Numerically they are
\begin{align}
    \Gamma_\text{lattice}(0.1) &= 2.09(4)\times 10^{-4} \,, \\
    \Gamma_\text{Langer}(0.1) &= 2.25(23)\times 10^{-4} \,,
\end{align}
where the lattice error is statistical and the perturbative error is due to two-loop corrections.
Discretization errors are expected to be of order 0.5\%, for which see Appendix~\ref{app:numericalTests}, and hence much smaller than statistical errors.

The field theory we have simulated has fluctuations at the lattice scale, so that taking the continuum limit requires some care.
Fortunately, this theory is superrenormalizable, and hence quickly approaches a free theory at high energies.
More specifically, the only divergent counterterm is a constant addition to the Hamiltonian, which affects neither the dynamics nor the statistics of the system; see Appendix~\ref{app:continuumLimit}.

Statistical errors on the lattice results are computed by multiplying the standard deviation by the square root of twice the integrated autocorrelation time of the observable under consideration. As expected, the computed errors on the binned nucleation rate are comparable to the statistical fluctuations between bins.

Results at all five temperatures simulated are arrayed in Fig.~\ref{fig:rates_compared}.
The time dependence of the rate is similar across temperatures, and in each case we can safely extrapolate in time $t\to 0$ to obtain the final rates.
Crucially, the results show that the lattice nucleation rate is tracked by Langer's formula.

To assess whether the lattice data aligns with expected higher-order corrections, we compare two models for the $\hat{T}$-dependence: (i) one including two-loop corrections, $\Gamma = \Gamma_\text{Langer}(1 + c_2\hat{T})$, and (ii) a simple rescaling, $\Gamma = c\Gamma_\text{Langer}$. Fitting the data yields $c_2 = -0.94(4)$ with $\chi^2/\text{d.o.f.} = 1.1$, and $c = 0.894(6)$ with $\chi^2/\text{d.o.f.} = 1.8$, indicating a marginally better fit for the perturbative model. A Bayesian analysis supports this, using priors for $c_2$ that are approximately uniform for $O(1)$ values and suppressed for larger magnitudes. We compare this to two alternative models based on the hypothesis that Langer's rate is significantly modified: one with an $O(1)$ multiplicative factor (log-normal prior) and one with an $O(10)$ factor (log-uniform prior); see Appendix~\ref{app:priors}. The resulting Bayes factors, 4.0 and 9.5 respectively, provide substantial to strong support for the perturbative model.
This stems from philosophical differences between the models; the perturbative model assigns more prior weight in near agreement with Langer's rate.

Fig.~\ref{fig:rate_histogram_gaussian_T0.1} shows that Gaussian initialization leads to a strongly time-dependent rate.
The rate starts from zero, because the Gaussian initialization gives exponentially smaller weight to configurations which are near the edge of the metastable phase, where $\frac{1}{2}m_T^2\phi^2 > V(\phi)$.
The rate then peaks around $t\approx 2.5$, corresponding to configurations that escape in a single motion.
Subsequent oscillations reflect configurations that bounce back and escape in later cycles.
All of these time-dependent features thwart extracting a precise nucleation rate for such initial conditions.

One might hope that after some initial transitory behaviour, the rate of nucleation would settle down to an approximately constant value.
This has been investigated in Refs.~\cite{Pirvu:2024nbe, Pirvu:2024ova}, in which they showed that over long times the rate decreases and remains significantly below Langer's formula at this parameter point.
Fig.~7 in Ref.~\cite{Pirvu:2024nbe} shows that the rate continues to decrease over times $O(10^3)$.

The rate shown in Fig.~\ref{fig:rate_histogram_T0.1} is also time dependent, decreasing by a factor half over the time $\mathcal{T}=10$.
However, unlike the rate following Gaussian initialization, it has a well-defined, nonzero limit for $t\to 0$, allowing us to extract the thermal rate following Eq.~\eqref{eq:rate_definition}.

\sectioninline{Conclusions}
For the first time, we have demonstrated the quantitative correctness of Eq.~\eqref{eq:langer_rate} for the nucleation rate in lattice simulations.
The discrepancy is dramatically smaller than found in all previous lattice studies, the closest of which found a multiplicative discrepancy of $\sim 8$ for the same parameter point~\cite{Pirvu:2024nbe, Pirvu:2024ova},
though using a different definition of the nucleation rate.
Most previous studies in other models have found multiplicative discrepancies between $e^{O(10)}$ and $e^{O(100)}$.

The precise agreement found between lattice and perturbation theory puts homogeneous nucleation theory on firm footing, and excludes
the mooted possibility of the failure of Langer's theory to describe homogeneous nucleation.
Thus, some other reason must explain the many orders of mangitude discrepancy between slower semiclassical and faster measured nucleation rates in ${}^3$He.

The residual 10\% difference between the lattice and Langer's semiclassical formalism is well explained by two-loop corrections, including the expected functional form.
Thus, a two-loop computation of the nucleation rate, following Ref.~\cite{Ekstedt:2022tqk}, see also \cite{Carosi:2024lop},
would extend the test of agreement down to 1\% and include new physical effects, such as curvature of the boundary between phases and back-reaction of the bubble on the thermal fluctuations.

A key development of this work has been a precise nonperturbative definition of the nucleation rate, that can be implemented on the lattice.
This also reveals the \textit{meaning} of nucleation implied by the perturbative expression.
The rate depends sensitively on the initial distribution, and hence in turn relies on a precise definition of the metastable phase.
This has important implications for simulations based on the truncated-Wigner approximation~\cite{Braden:2018tky, Billam:2018pvp, Hertzberg:2020tqa},
that only accounts for Gaussian initial fluctuations, but also for analog experiments which will require a precise initial state.

In agreement with previous works, we find that observations of a time-dependent nucleation rate are a consequence of incomplete thermalization of the full metastable phase.
This we confirmed by factorizing the simulation into first thermalisation, and then conservative evolution.
A completely thermalised metastable phase nucleates with Langer's rate, as we have demonstrated.

However, in a dynamical scenario, will the system thermalise sufficiently quickly to nucleate at Langer's rate?
It has been argued that this occurs only if~\cite{Hanggi:1990zz}
\begin{equation} \label{eq:nucleation_thermalisation}
    t_\text{th} \sqrt{|\lambda_-|} < \beta H[\phi_\text{cb}] \,,
\end{equation}
in terms of the thermalization time $t_\text{th}$, and the negative eigenvalue $\lambda_-$ of fluctuations around the critical bubble.
When this condition fails, configurations near the critical bubble are unable to thermalise.

For the parameter points we have studied, the observed decrease of the nucleation rate over relatively short times suggests that this condition is not met.
Thus, we expect Eq.~\eqref{eq:rate_definition} overestimates the time-dependent rate obtained in a dynamical case.

More generally, the appropriate value for $\beta H[\phi_\text{cb}]$ depends on the temperature at which the system is transitioning, typically when an $O(1)$ fraction is in each phase.
This reflects both microscopic and macroscopic properties of the system, such as how quickly the temperature changes through the transition, with slower changes leading to transitions at larger $\beta H[\phi_\text{cb}]$.
By contrast, the left-hand side of Eq.~\eqref{eq:nucleation_thermalisation} is determined purely by microphysics, being larger for more weakly coupled fields.

In Big Bang cosmology, the slow cooling due to Hubble expansion leads to large values of $\beta H[\phi_\text{cb}]$, growing with $\sim 4\log(M_\text{Planck}/T)$.
For instance $\beta H[\phi_\text{cb}]\sim 40$ at grand unification, $\sim 140$ at the electroweak scale, and $\sim 170$ at the QCD scale~\cite{Enqvist:1991xw}.
A Boltzmann analysis predicts the thermalization time is approximately $t_\text{th}\sim 1/(\alpha^2 T)$ for a relativistic plasma in 3+1 dimensions, with $\alpha$ the relevant fine-structure constant~\cite{Kurkela:2011ti}.%
\footnote{
The Boltzmann analysis parametrically underestimates the thermalization of the bubbles if they are coupled to bosonic fields lighter than the temperature~\cite{Hirvonen:2024rfg}.
}

Putting this together for a first-order transition at the electroweak scale, and assuming a one-step transition with $\sqrt{|\lambda_-|}\sim \alpha^{3/4}T$~\cite{Ekstedt:2022zro}, suggests that models with couplings $\alpha \gtrsim 1/50$ should thermalise, and hence nucleate with Langer's rate.
More weakly coupled fields will fail to thermalize, and hence will nucleate more slowly.
However, this point deserves further scrutiny.

Finally, a key direction for future study is the generalization to higher dimensions, particularly in light of Ref.~\cite{Gould:2024chm}, which found much larger discrepancies between lattice and perturbative nucleation rates in 3+1 dimensions.
In higher dimensions, effects from high energy thermal fluctuations are more pronounced, but even in 3+1 dimensions these are superrenormalizable and hence tractable.
Thus, extending our approach to higher dimensions offers clear promise.


\begin{acknowledgments}
    We would like to thank Andrey Shkerin and Sergey Sibiryakov for their insightful comments on the manuscript.
    J.H.\ (ORCID 0000-0002-5350-7556)
    and O.G.\ (ORCID ID 0000-0002-7815-3379)
    were supported by a Royal Society Dorothy Hodgkin Fellowship.
    We are grateful for access to the University of Nottingham's Ada HPC service.
\end{acknowledgments}

\section*{Data availability statement}
Data shown in Figs.~\ref{fig:rate_histogram_T0.1}, \ref{fig:rates_compared} and \ref{fig:rate_histogram_gaussian_T0.1} in this paper are available at Ref.~\cite{hirvonen_2025_15424939}.

 \newcommand{\noop}[1]{}
%



\clearpage

\begin{widetext}
\begin{center}

  {\large \bf Supplemental material}

\end{center}
\end{widetext}

\setcounter{section}{0}


\section{Lattice details}\label{app:LatticeDetails}

Here, we will detail our lattice implementation of the model. We begin with the spatially discretized Hamiltonian and discuss the spatial discretization being sufficiently close to the continuum for reproducing Langer's rate. Then, we continue with the time-evolution algorithm, which is symplectic, fourth-order algorithm.

The Hamiltonian is given by
\begin{align}
    H[\phi,\pi]&=\sum_{n=1}^{\Nsites}\delta x \qty(\frac{1}{2}\pi_n^2-\frac{1}{2}\phi_n(\partial^2\phi)_n+V(\phi_n)) \,,\\
    (\partial^2\phi)_n&=\frac{\phi_{n-1}+\phi_{n+1}-2\phi_n}{\delta x^2} \,,\\
    V(\phi)&=\frac{1}{2}\phi^2-\frac{1}{4}\phi^4 \,.
\end{align}
Here, $\delta x$ is the distance between lattice sites, and the boundary conditions are periodic: $(\phi_{\Nsites+1}, \pi_{\Nsites+1})\equiv(\phi_1, \pi_1)$, $(\phi_{-1}, \pi_{-1})\equiv(\phi_{\Nsites}, \pi_{\Nsites})$.

We used the Forest--Ruth algorithm~\cite{Forest:1989ez} to implement time evolution for the system. It is a four-order accurate, symplectic integrator that can be constructed from the second-order accurate leap-frog algorithm~\cite{Yoshida:1990zz}. A single leap-frog step is given by the following algorithm:
\begin{align}
    \pi_n\qty(t+\frac{\delta t}{2})&=\pi_n(t)-\frac{1}{\delta x}\pdv{H}{\phi_n}[\phi(t)]\,\frac{\delta t}{2} \,,\\
    \phi_n(t+\delta t)&=\phi_n(t)+\pi_n\qty(t+\frac{\delta t}{2})\,\delta t \,,\\
    \pi_n(t+\delta t)&=\pi_n\qty(t+\frac{\delta t}{2})-\frac{1}{\delta x}\pdv{H}{\phi_n}[\phi(t+\delta t)]\,\frac{\delta t}{2} \,.
\end{align}
The Forest--Ruth step is then given by performing the leap-frog step with time steps $\delta t_1, \delta t_2, \delta t_3$,
\begin{align}
    \delta t_1&=(2-2^{1/3})^{-1} \delta t \,,\\
    \delta t_2&=-2^{1/3}(2-2^{1/3})^{-1} \delta t \,,\\
    \delta t_3&=\delta t_1 \,,
\end{align}
where $\delta t$ gives the total change in physical time for a single Forest--Ruth step.

\section{Continuum limit}\label{app:continuumLimit}

The lattice field configurations sampled from the Boltzmann distribution are not smooth, and hence the question of the continuum limit requires special consideration.

The field theory we consider has a single coupling $\lambda$ with mass dimension 2. In the language of the renormalization group, the corresponding operator is relevant in the infrared, and conversely irrelevant in the ultraviolet. In other words, the theory is super-renormalizable.

The approach to the continuum limit can be determined by considering the effective potential within lattice perturbation theory~\cite{Laine:1995np}.
Taking first the infinite volume limit but retaining the dependence on the lattice spacing, the effective potential at one-loop is~\cite{Pirvu:2024nbe}
\begin{align} \label{eq:effective_potential_lattice_one_loop}
    V_\text{eff}^{(1)}(\phi) &= \frac{T}{2}\int_{-\frac{\pi}{\delta x}}^{\frac{\pi}{\delta x}} \frac{\mathrm{d}p}{2\pi}\log\left[\frac{\sin^2\left(\frac{p\delta x}{2}\right)}{(\delta x / 2)^2} + m^2 - 3 \lambda \phi^2\right] \,,
    \nonumber \\
    &= -\frac{T\log\delta x}{\delta x} + \frac{T}{2}\sqrt{m^2 - 3\lambda \phi^2}  + O(\delta x^2)\,.
\end{align}
This expression contains a divergent constant term as $\delta x\to 0$, which can be canceled by a counterterm for the cosmological constant. Regardless, it affects neither the dynamics nor the statistics of the model. There are no ultraviolet divergences which have field dependence, thus the Lagrangian parameters are not renormalized. Due to the positive mass dimension of the coupling $\lambda$, higher-loop contributions are even less sensitive to the ultraviolet cut off.

Note that the full quantum field theory does contain field-dependent ultraviolet divergences, in particular for the mass term. This can be canceled in the usual way with appropriate counterterms~\cite{Pirvu:2024nbe}. However, after matching to the classical theory at finite temperature, the effective model parameters are finite.

\section{Thermalization algorithm}\label{app:Thermalization}

For thermalizing the metastable phase, we used a slightly modified version of Hamiltonian Monte Carlo, also known as Hybrid Monte Carlo~\cite{Duane:1987de}.
The sampled distribution function corresponds to equilibrium in the metastable phase:
\begin{equation}\label{eq:momentumHeatBathMetastableThermaDistribution}
    \rho_\text{meta}[\phi]=
    \begin{cases}
        \frac{1}{Z_{\phi, \text{meta}}}e^{-\beta \Phi[\phi]} \,, & \phi\in\text{meta phase}\\
        0 \,, & \phi\notin\text{meta phase}
    \end{cases}\,.
\end{equation}
The weight is given by the field-dependent part of the Hamiltonian,
\begin{equation}
    \Phi[\phi]=H[\phi,\pi=0] \,,
\end{equation}
and the partition function, $Z_{\phi, \text{meta}}$, normalizes the distribution to unity.
The configuration $\phi$ is in the metastable phase if the solution to the gradient descent,
\begin{equation}
    \dv{\phi}{\tau}=-\grad_\phi\Phi \,,\qquad\phi(\tau=0)=\phi \,,
\end{equation}
asymptotes to zero, $\phi(\tau\to\infty)\to0$.

One Monte Carlo step consists of $N$ substeps in which:
\begin{enumerate}
    \item Conjugate momenta are thermalized,
\begin{equation}\label{eq:drawingMomenta}
    \pi \sim \frac{e^{-\beta\Pi[\pi]}}{Z_\pi} \,,
\end{equation}\label{enum:momentumThermalization}
    \item The system is evolved for time $\Delta t$ with the integrator of Sec.~\ref{app:LatticeDetails}.\label{enum:hamiltonianEvolution}
\end{enumerate}
Here, the weight is given by the conjugate-momentum-dependent part of the Hamiltonian
\begin{equation}
    \Pi[\pi]=H[\phi=0,\pi] \,,
\end{equation}
and the partition function, $Z_{\pi}$, normalizes the distribution to unity. 

In order to sample correctly the distribution in Eq.~\eqref{eq:momentumHeatBathMetastableThermaDistribution}, the method must be ergodic and obey detailed balance. Neither of the steps inside a substep is ergodic on their own: step~\ref{enum:momentumThermalization} conserves $\phi$ and step~\ref{enum:hamiltonianEvolution} conserves the Hamiltonian of our integrator, Sec.~\ref{app:LatticeDetails}, that approximates the physical Hamiltonian~\cite{approximateHamiltonian}. However, together they are ergodic: In step~\ref{enum:momentumThermalization}, any conjugate momentum is possible. There are $\Nsites$ degrees of freedom in a conjugate momentum. There are the same number of degrees of freedom in the field. Thus, one should be able to find a conjugate momentum for getting from one configuration to any other. Hence, even a single substep is ergodic. The $N$ substeps only serve to diminish correlations between configurations in the chain, and to make the algorithm faster.

For the method to obey detailed balance, the $N$ substeps detailed above are not enough. The probability weight for jumping from configuration $\phi_1$ to $\phi_2$ in a single Monet Carlo step, $W(\phi_1\to\phi_2)$, must satisfy the following condition: 
\begin{align}\label{eq:detailedBalance}
    \rho_\text{meta}[\phi_1]\,W(\phi_1\to\phi_2)
    =\rho_\text{meta}[\phi_2]\, W(\phi_2\to\phi_1) \,.
\end{align}
For this to hold, we must account for the fact that the system may have evolved outside the metastable phase, $\rho_\text{meta}[\phi_\text{out}]=0$, and that the integrator does not exactly conserve the physical Hamiltonian.

The probability weight for the system to transfer outside the metastable phase in a total Monte Carlo step must be zero:
\begin{align}
    W(\phi\to\phi_\text{out})
    =\frac{\rho_\text{meta}[\phi_\text{out}]}{\rho_\text{meta}[\phi]}W(\phi_\text{out}\to\phi)=0 \,.
\end{align}
If the $N$ substeps have taken the system outside the phase, the ending point of the total step is to return to the previous configuration.

Correcting for the integrator not being exact requires a Metropolis--Hastings step~\cite{Metropolis:1953am,Hastings} that is extended to apply to $N$ substeps. Let us define a process $\gamma=(\pi_1,\pi_2,\dots,\pi_N)$, which contains all of the thermal conjugate momenta drawn from the distribution in Eq.~\eqref{eq:drawingMomenta}. In addition to the initial configuration, $\phi_1$, the process $\gamma$ contains all of the information regarding the $N$ substeps. Hence, the process unfolds as
\begin{align}
    \phi_1&\xrightarrow{\text{step~\ref{enum:momentumThermalization}}}(\phi_1,\pi_1)
    \xrightarrow{\text{step~\ref{enum:hamiltonianEvolution}}}(\phi_1',\pi_1')\\
    &\xrightarrow{\text{step~\ref{enum:momentumThermalization}}}(\phi_2,\pi_2)
    \xrightarrow{\text{step~\ref{enum:hamiltonianEvolution}}}\dots\\
    &\xrightarrow{\text{step~\ref{enum:momentumThermalization}}}(\phi_N,\pi_N)
    \xrightarrow{\text{step~\ref{enum:hamiltonianEvolution}}}(\phi_N',\pi_N') \,,
\end{align}
where $\phi_i'=\phi_{i+1}$, and $\phi_N'$ is the field configuration resulting from the $N$ substeps.

Notice that for every process $\gamma$ that takes the configuration from $\phi_1$ to $\phi_N'$ there is a counterprocess $-\gamma'=(-\pi_N',-\pi_{N-1}',\dots,-\pi_1')$ from $\phi_N'$ to $\phi_1$. This is due to the time reversibility of the system and the integrator.

It will be sufficient for detailed balance, Eq.~\eqref{eq:detailedBalance}, to enforce every process and its counterprocess to obey detailed balance:
\begin{align}\label{eq:processDetailedBalance}
    &\quad\;\rho_\text{meta}[\phi_1]\,W(\phi_1\xrightarrow{\gamma}\phi_N')\nonumber\\
    &=\rho_\text{meta}[\phi_N']\, W(\phi_N'\xrightarrow{-\gamma'}\phi_1) \,.
\end{align}
We can integrate over the space of possible processes from $\phi_1$ to $\phi_N'$, $\gamma$, and the counterprocesses, $-\gamma'$, symmetrically on their respective sides of the equation to obtain the condition for total detailed balance, Eq.~\eqref{eq:detailedBalance}.

The ``raw'' probability weight of the process $\gamma$, which corresponds to performing only the $N$ substeps, is given by
\begin{equation}
    \overline{W}(\phi_1\stackrel{\gamma}{\to}\phi_N')=\prod_{i=1}^N \frac{e^{-\beta\Pi[\pi_i]}}{Z_\pi} \,.
\end{equation}
We can extract the difference of the process and its counterprocess in the detailed-balance condition in Eq.~\eqref{eq:processDetailedBalance},
\begin{align}
    &\quad\;\rho_\text{meta}[\phi_1]\,\overline{W}(\phi_1\xrightarrow{\gamma}\phi_N')\label{eq:startingPointRawProcess}\\
    &=\frac{e^{-\beta \Phi[\phi_1]}}{Z_{\phi, \text{meta}}}\prod_{i=1}^N \frac{e^{-\beta\Pi[\pi_i]}}{Z_\pi}\\
    &=\frac{e^{-\beta H[\phi_1,\pi_1]}}{Z_{\phi, \text{meta}}Z_\pi}\prod_{i=2}^N \frac{e^{-\beta\Pi[\pi_i]}}{Z_\pi}\\
    &=e^{-\beta \Delta \!H[\phi_1,\pi_1]} \frac{e^{-\beta H[\phi_1',\pi_1']}}{Z_{\phi, \text{meta}}Z_\pi}\prod_{i=2}^N \frac{e^{-\beta\Pi[\pi_i]}}{Z_\pi}\\
    &=e^{-\beta \Delta \!H[\phi_1,\pi_1]} \frac{e^{-\beta \Pi[\pi_1']}}{Z_\pi}\frac{e^{-\beta \Phi[\phi_2]}}{Z_{\phi, \text{meta}}}\prod_{i=2}^N \frac{e^{-\beta\Pi[\pi_i]}}{Z_\pi}\\
    &=\frac{e^{-\beta \Phi[\phi_N']}}{Z_{\phi, \text{meta}}}\prod_{i=1}^N e^{-\beta \Delta \!H[\phi_i,\pi_i]} \frac{e^{-\beta\Pi[\pi_i']}}{Z_\pi}\\
    &=\rho_\text{meta}[\phi_N']\,\overline{W} (\phi_N'\xrightarrow{-\gamma'}\phi_1)\prod_{i=1}^N e^{-\beta \Delta \!H[\phi_i,\pi_i]} \,.\label{eq:endingPointRawProcess}
\end{align}
Here, we have defined $\Delta \!H[\phi_i,\pi_i]\equiv H[\phi_i',\pi_i']-H[\phi_i,\pi_i]$, which is in general non zero due to the non-exact integrator. The difference is given by the changes in the Hamiltonian during the time evolution, $\Delta \!H[\phi_i,\pi_i]$.

For a simple Metropolis--Hastings step to correct the bias found above, the bias must be the same for an antiprocess as well. This is ensured by the time reversibility. We can use the above equality between Eqs.~\eqref{eq:startingPointRawProcess} and \eqref{eq:endingPointRawProcess} for a process on an antiprocess:
\begin{align}
    &\quad\;\rho_\text{meta}[\phi_N']\,\overline{W} (\phi_N'\xrightarrow{-\gamma'}\phi_1)\\
    &= \rho_\text{meta}[\phi_1]\,\overline{W}(\phi_1\xrightarrow{\gamma}\phi_N')\prod_{i=1}^N e^{-\beta \Delta \!H[\phi_i',-\pi_i']}\\
    &=\rho_\text{meta}[\phi_1]\,\overline{W}(\phi_1\xrightarrow{\gamma}\phi_N')\prod_{i=1}^N e^{+\beta \Delta \!H[\phi_i,\pi_i]} \,.
\end{align}
The latter equality follows from the time reversibility: $\Delta \!H[\phi_i',-\pi_i']=H[\phi_i,\pi_i]-H[\phi_i',\pi_i']=-\Delta \!H[\phi_i,\pi_i]$. Notice that the equality between the first and the last line is the equality between Eqs.~\eqref{eq:startingPointRawProcess} and \eqref{eq:endingPointRawProcess}, showing that the bias is consistent.

The process and its counterprocess can be made to obey detailed balance by accepting the new configuration with probability of $\min(1, e^{-\beta \sum_i \Delta \!H[\phi_i,\pi_i]})$. After adding the accept-reject step, the probability weight for a process is
\begin{align}
\begin{split}
    W(\phi_1\xrightarrow{\gamma}\phi_N')&=\overline{W}(\phi_1\xrightarrow{\gamma}\phi_N')\\
    &\quad\;\times\min(1, e^{-\beta \sum_i \Delta \!H[\phi_i,\pi_i]}) \,,
\end{split}
\end{align}
which obey detailed balance of process and its counterprocess, Eq.~\eqref{eq:processDetailedBalance}.

\begin{figure}[t]
    \includegraphics*[width=0.95\columnwidth]{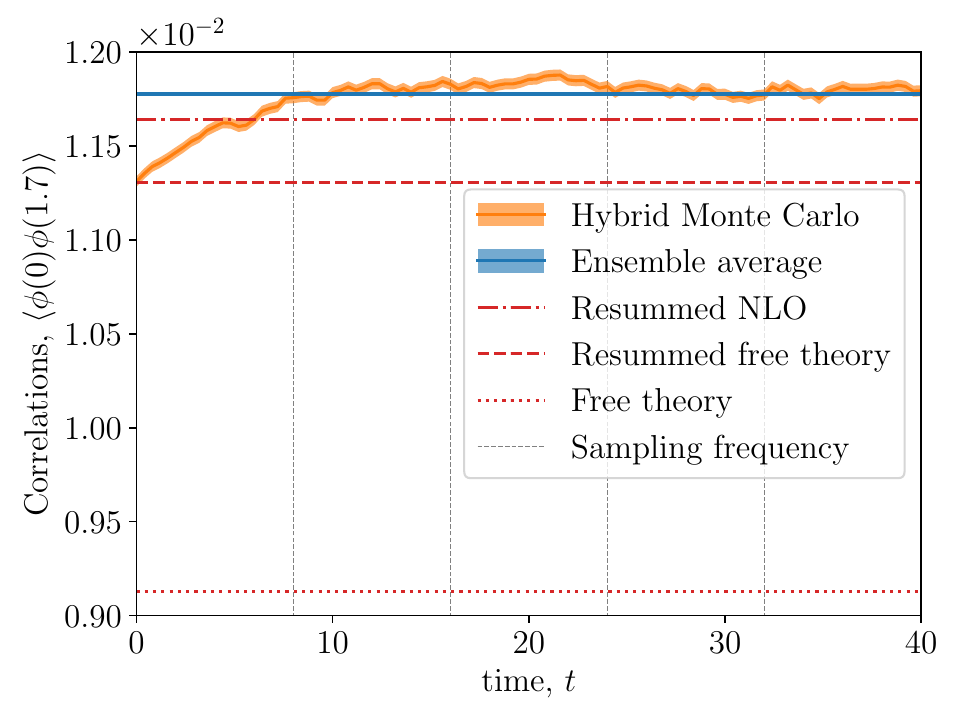}
    \caption{The two-point correlator $\left< \phi(0) \phi(1.7) \right>$ upon initialization and subsequent ``burn in'' of the hybrid Monte-Carlo simulation. At $t=0$ the configuration is initialized as a random free field with the one-loop resummed mass, and then evolves nonperturbatively according to the hybrid Monte-Carlo algorithm. Samples are taken periodically at the sampling frequency, with the first 4000 being discarded.}
    \label{fig:burn_in}
  \end{figure}

\section{Numerical tests}\label{app:numericalTests}

Here we report on the tests performed on the code. For our chosen parameter points, the expected two-loop uncertainty is of order 10\%, hence we aim to keep lattice effects significantly below this level.

We chose the time step to be $\delta t=0.02$, which conserves the energy to a relative order of $10^{-4}$.
We also checked that the algorithm behaves as expected:
It is of fourth order in $\delta t/\delta x$,
the energy is conserved to the fifth order,
and the time step is small enough for the ultraviolet modes of the system to be treated accurately, given $\delta x=0.1$.

We carry out a direct test of the size of lattice effects for the nucleation rate, by computing Langer's rate both in the continuum and with the discretized Hamiltonian. Note these are both perturbative, one-loop computations. The continuum result is given in Eq.~\eqref{eq:langer_in_our_model} and the latter is
\begin{equation} \label{eq:langer_discrete}
    \Gamma_{\text{Langer}}^{\delta x}=\frac{\Nsites\delta x}{2\pi}
    \left(\frac{H_\text{cb}}{2\pi T}\frac{\prod_{n=1}^{\Nsites}\lambda^\text{meta}_n}{\prod_{n=3}^{\Nsites}\lambda^\text{cb}_n}\right)^{\frac{1}{2}} e^{-H_\text{cb}/T}.
\end{equation}
The Hamiltonian is evaluated on the critical bubble, $H_\text{cb}=H[\cb,0]$, where $\cb$ is the critical bubble of the discretized Hamiltonian. The eigenvalues, $\lambda$, are the eigenvalues of the field dependent Hamiltonian expanded around the metastable phase, $H[0+\Delta\phi,0]\approx \frac{1}{2}\Delta\phi\cdot H^{(2)}[0,0]\cdot \Delta\phi$, and the critical bubble, $H[\cb+\Delta\phi,0]\approx H_\text{cb} + \frac{1}{2}\Delta\phi\cdot H^{(2)}[\cb,0]\cdot \Delta\phi$. For the latter set, the negative and translational eigenmodes are not included in the product.

With our choice of lattice spacing, $\delta x=0.1$, the discrete and continuum expressions for Langer's rate, Eqs.~\eqref{eq:langer_discrete} and \eqref{eq:langer_in_our_model}, agree to within 0.5\% for each of the temperatures simulated.

One subtlety is that the above discrete Langer's rate assumes translational invariance, which is broken by lattice discretization. We also verified this to be a completely negligible error by computing the discrete rate with two distinct critical bubbles whose centers were at a lattice site and in the middle of two sites. To give a sense of how minutely the translational invariance is broken, the magnitude of the translational eigenmode was below $10^{-11}$, whereas the other two of the three lowest eigenvalues were -3 and 0.1.

We have also tested agreement with perturbation theory for the connected two-point function of our theory, on a distance scale comparable to the critical bubble radius $|x-x'|=1.7$ (specifically the distance between the inflection points of the leading-order critical bubble).
This is reported in Fig.~\ref{fig:burn_in}.

We compare to both vanilla and resummed perturbation theory. The two-point function of the resummed free-theory is
\begin{equation} \label{eq:resummed_free_twopoint}
    \langle \delta \phi(x) \delta \phi(x')\rangle_{0} = \frac{\hat{T}}{2m_T}e^{-m_T|x-x'|} \,.
\end{equation}
The unresummed free two-point function has simply $m_T\to m$.

\begin{figure}[t]
    \subfloat[~]{
        \includegraphics{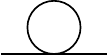}
    }\hfill%
    \subfloat[~]{
        \includegraphics{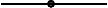}
    }\hfill%
    \subfloat[~]{
        \includegraphics{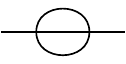}
    }\hfill
    \caption{The Feynman diagrams that contribute to the two-point function, $\left<\phi(x)\phi(x')\right>$, at next-to-next-to-leading order in perturbation theory, after mass resummation. Here the blob is a mass counterterm which avoids double counting. All other one-loop and two-loop diagrams cancel up to higher powers of $\hat{T}$, due to the resummation. The result is given in Eq.~\eqref{eq:twopoint}.}
    \label{fig:feynman-diagrams}
\end{figure}

Our hybrid Monte-Carlo simulation is initialized to the resummed free theory. Fig.~\ref{fig:burn_in} shows perfect agreement between the lattice data and Eq.~\eqref{eq:resummed_free_twopoint} at $t=0$.
Then, for $t>0$ the system tends towards larger values of the two-point function under the non-perturbative evolution. This too is in agreement with expected higher order corrections,
\begin{align} \label{eq:twopoint}
    &\frac{\langle \delta \phi(x) \delta \phi(x')\rangle}{\langle \delta \phi(x) \delta \phi(x')\rangle_0} =  
    1 +
    \frac{3\hat{T}(m-m_T)}{4 m_T^3m}
    \left(1 + m_T|x-x'| \right)
    \nonumber\\
    &\quad
    +\frac{3\hat{T}^2}{128 m_T^6}
    \left(9 + 12 m_T |x-x'| + e^{-2m_T|x-x'|}
    \right)
    \,,
\end{align}
up to $O(\hat{T}^3)$ corrections.
In computing these higher order terms, we have carried out an explicit resummation by adding a mass correction to the free Lagrangian, and taking it away from the interacting Lagrangian.
Combinations of diagrams which together contribute at higher orders after cancellations have been dropped.
The relevant Feynman diagrams are shown in Fig.~\ref{fig:feynman-diagrams}.

Further, we have repeated the calculation of Eq.~\eqref{eq:twopoint} in lattice perturbation theory for our lattice parameters, and find agreement with the continuum perturbative result to 0.07\%.

We chose the lattice size to be $L=50$, at which point it is $\sim 46$ times the Compton wavelength, and $\sim 15$ times the critical bubble diameter. Finite size corrections to the critical bubble decrease exponentially $\propto e^{-m_T L}$ for large lattices, so should be negligible. Corrections to the thermal mass $\delta m_T^2\sim \hat{T}/L$ are at the level of 0.2\%. Ref.~\cite{Pirvu:2024nbe} found that the nucleation rate per unit volume agreed within statistical uncertainties between $L=50, 100, 200$.

\section{Priors for model fits} \label{app:priors}

Our priors for the perturbative model $\Gamma=\Gamma_\text{Langer}(1+c_2\hat{T})$ were chosen to be approximately uniform over $O(1)$ values of $c_2$, and decreasing exponentially for larger magnitudes,
\begin{equation}
    p_\text{prior}^{\text{perturbative}}(c_2) = \mathcal{N} \exp\left(-\frac{(A^4 + c_2^4)^{1/4}}{A}\right) \,,
\end{equation}
with $A=\sqrt{10}$, and $\mathcal{N}$ a normalization constant.

Our priors for the models assuming Langer's rate is modified multiplicatively, $\Gamma=c \Gamma_\text{Langer}$, were chosen to be symmetric in log space around $x=1$, so that $c=1/x$ is equally probable as $c=x$.
In particular, for the model assuming an $O(1)$ deviation we chose
\begin{equation}
    p^{O(1)}_\text{prior}(c) = \mathcal{N}' c^{-1} \exp\left(-\frac{\log(c)^2}{2\sigma^2}\right) \,,
\end{equation}
with $\sigma=0.7$ and $\mathcal{N}'$ a normalization constant. This value for $\sigma$ leads to 90\% of the prior weight being in the range $c\in[1/\sqrt{10}, \sqrt{10}]$.

For the model assuming an $O(10)$ deviation from Langer's rate we chose an approximately log-uniform prior in the range $c\in[1/10, 10]$,
\begin{equation}
    p^{O(10)}_\text{prior}(c) = \frac{\mathcal{N}''}{\frac{B^2}{c} + B + c + B c^2 + B^2 c^3} \,,
\end{equation}
where $B=1/10$ and $\mathcal{N}''$ is a normalization constant. 


\end{document}